\documentclass[10pt,conference]{IEEEtran}\IEEEoverridecommandlockouts
\usepackage{cite, url}
\usepackage{amsmath,amssymb,amsfonts}
\usepackage{algorithmic}
\usepackage{graphicx}
\usepackage{textcomp}
\usepackage{xcolor}
\usepackage{hyperref}
\usepackage{multirow}
\usepackage{tcolorbox}
\definecolor{darkgreen}{rgb}{0.0, 0.5, 0.0}
\newtcolorbox{findingsbox}[1][]{
  colback=gray!10,    
  colframe=black,     
  fonttitle=\bfseries, 
  title=#1,           
  coltitle=black,     
  boxrule=0.5pt,      
  left=1mm,           
  right=1mm,          
  top=1mm,            
  bottom=1mm          
}
\def\BibTeX{{\rm B\kern-.05em{\sc i\kern-.025em b}\ker 
 n-.08em
    T\kern-.1667em\lower.7ex\hbox{E}\kern-.125emX}}
\begin{document}

\title{Prompting and Fine-tuning Large Language Models for Automated Code Review Comment Generation
\thanks{* All three authors have equal contributions.}\\
\thanks{The funding from the Research and Innovation Centre For Science and Engineering, RISE-BUET Undergraduate Student Research Grant S2023-03-113 supports our work.}
}

\author{Anonymous Author(s)}

\author{\IEEEauthorblockN{
Md. Asif Haider\textsuperscript{1*}, 
Ayesha Binte Mostofa\textsuperscript{1*}, 
Sk. Sabit Bin Mosaddek\textsuperscript{1*}, 
Anindya Iqbal\textsuperscript{1}, 
Toufique Ahmed\textsuperscript{2}
}
\IEEEauthorblockA{
\textsuperscript{1}\textit{Bangladesh University of Engineering and Technology}, Dhaka, Bangladesh \\
\textsuperscript{2}\textit{University of California}, Davis, USA \\
1805112@ugrad.cse.buet.ac.bd, 1805062@ugrad.cse.buet.ac.bd, 1805106@ugrad.cse.buet.ac.bd, \\ anindya@cse.buet.ac.bd, 
tfahmed@ucdavis.edu
}
}
\maketitle
\bibliographystyle{IEEEtran}
\begin{abstract}
Generating accurate code review comments remains a significant challenge due to the inherently diverse and non-unique nature of the task output. Large language models pretrained on both programming and natural language data tend to perform well in code-oriented tasks. However, large-scale pretraining is not always feasible due to its environmental impact and project-specific generalizability issues. In this work, first we fine-tune open-source Large language models (LLM) in parameter-efficient, quantized low-rank (QLoRA) fashion on consumer grade hardware to improve review comment generation. Recent studies demonstrate the efficacy of augmenting semantic metadata information into prompts to boost performance in other code-related tasks. To explore this in code review activities, we also prompt proprietary, closed-source LLMs augmenting the input code patch with function call graphs and code summaries. Both of our strategies improve the review comment generation performance, with function call graph augmented few-shot prompting on the GPT-3.5 model surpassing the pretrained baseline by around \textbf{$90\%$} BLEU-4 score on the CodeReviewer dataset. Moreover, few-shot prompted Gemini-1.0 Pro, QLoRA fine-tuned Code Llama and Llama 3.1 models achieve competitive results (ranging from $25\%$ to $83\%$ performance improvement) on this task. An additional human evaluation study further validates our experimental findings, reflecting real-world developers' perceptions of LLM-generated code review comments based on relevant qualitative metrics. 
\end{abstract}


\section{Introduction}

Code review, the manual process of inspecting authored source code by fellow teammates, is a crucial part of the software development lifecycle that helps detect errors and encourages further code improvement \cite{spadini2020primers}.  First formalized by Fagan \cite{fagan2011design}, it is a systematic and collaborative software quality assurance activity where developers check each other’s code for improvement. Code reviews not only help in identifying bugs and potential issues early in the development cycle but also enhance code readability, maintainability, and overall software quality. However, despite the numerous benefits, the traditional review process demands significant manual effort, forcing developers to spend an excessive amount of time (more than 6 hours per week) reviewing their peers' code, as shown in \cite{bosu_review_cost, yang2016mining}. It is also responsible for frequent context switch of the developers from the actual tasks they are expected to focus on\cite{czerwonka2015code}. Hence, automating code review activities is in significant demand. One distinct task stands out in the Modern Code Review (MCR) practices: \textbf{Review Comment Generation (RCG)}, which can help reduce the burden from a code reviewer, automatically suggesting a potential change in the code submitted for review. We focus on improving the automation performance of review comment generation task in this study. \\



With the rapid advances in deep learning and natural language processing techniques, researchers proposed many Pretrained Language Models (PLM) on source code focusing on review tasks, notably including encoder-only BERT models \cite{feng2020codebert, guo2020graphcodebert} and encoder-decoder based T5 models \cite{wang2021codet5, li2022automating, tufano2022using}. Novel pretraining and fine-tuning attempts on large-scale datasets showed promising results. However, training on such domain-specific huge datasets requires a substantial amount of costly computing resources, imposing a negative impact on the carbon footprint globally \cite{strubell2020energy}.  While these models can usually generalize well, they might lack deep knowledge of project specific codebases, organizational coding standards, or niche libraries. This can lead to generic or less relevant code review suggestions, missing context-specific nuances.\\

However, decoder-only unified language models (e.g. based on GPT architecture) have shown superior performance when scaled to large parameter sizes. Also generally known as LLMs, these models can reduce the need for repetitive training while offering amazing few-shot learning capabilities \cite{brown2020language}. This refers to prompt engineering of the model with a few similar query-response pairs, also known as \textit{Few Shot Learning}. Designing efficient LLM prompts for the mentioned task yet remains less explored, motivating us toward this research direction.\\

Apart from proprietary LLMs, there has been a lot of work going on in the open-source landscape. General purpose open-source LLMs (e.g. Llama, Mistral) when fine-tuned, show performance improvement over PLMs. LLMs further trained on code-specific data, also known as Code LLMs (e.g. Codex, Code Llama) are currently the superior options for various code-related subtasks (including code generation, code summarization) \cite{zheng2023survey}. The best-performing versions of these LLMs nearly have 30-70B parameters, which is quite impossible to fit into a consumer grade GPU having around 16GB VRAM. Hence, fine-tuning the smaller versions of these LLMs (7-8B) is a promising cost-effective strategy to ensure project-specific, context-aware use cases. \textit{Parameter Efficient Fine-Tuning (PEFT)} approaches \textit{(Low-Rank Adaptation, 4-bit Quantization)} are found to assist in such endeavors \cite{lu2023llama}. \\

Augmenting statically analyzed, semantic structural facts to prompt the code model proved to be beneficial in code summarization tasks \cite{ahmed2024automatic}. Inspired by this, we propose a new methodology to design cost-effective few-shot prompts for proprietary LLMs, augmented with a programming language component- \textit{function call graph} and a natural language component- \textit{code summary}. We also explore further ablation studies to understand their standalone contributions for code review comment generation task. Additionally, we fine-tune open-source general-purpose LLMs and code LLMs in automating review comment generation task in a low resource-constrained setting. Specifically, we leverage the Quantized Low-Rank Adaptation (QLoRA) approach to fine-tune our models in a supervised way. To summarize, we particularly investigate the following research questions in this study:\\


\textbf{RQ1:} How effective is code review comment generation using fine-tuned open-source Large Language Models?

\textbf{RQ2:} How well do the closed-source Large Language Models perform in code review comment generation task when prompt engineered in a few-shot setting?

\textbf{RQ3:} When incorporated into prompts, what are the impacts of function call graph and code summary in improving review comment generation performance?

\textbf{RQ4:} How effective Large Language Models are in generating review comments from a real-world developer's perspective?

Our contributions can be summarized as follow:\\

\begin{itemize}
    \item Evaluating code review comment generation performance with open-source LLMs (Llama 2, Code Llama, Llama 3 series) in quantized, low-rank adapted parameter-efficient fine-tuning setup \\

    \item Exploring the performance of different closed-source, proprietary LLMs (GPT 3.5-Turbo, GPT-4o, Gemini-1.0 Pro) in few-shot prompt setting without any further data augmentation \\

    \item Investigating the impact of incorporating manually extracted function call graph and CodeT5 model generated code summary into the few-shot prompts to observe their impact on review comment generation performance \\

    \item Manual analysis and a developer study focusing on evaluating the LLM-generated review comments based on relevant qualitative metrics \\

    \item A replication package with all the scripts for data, code and result processing for our study, which can be found \href{https://doi.org/10.5281/zenodo.14063758}{here}.
\end{itemize}

\begin{figure*}[htbp]
\centerline{\includegraphics[width=\textwidth]{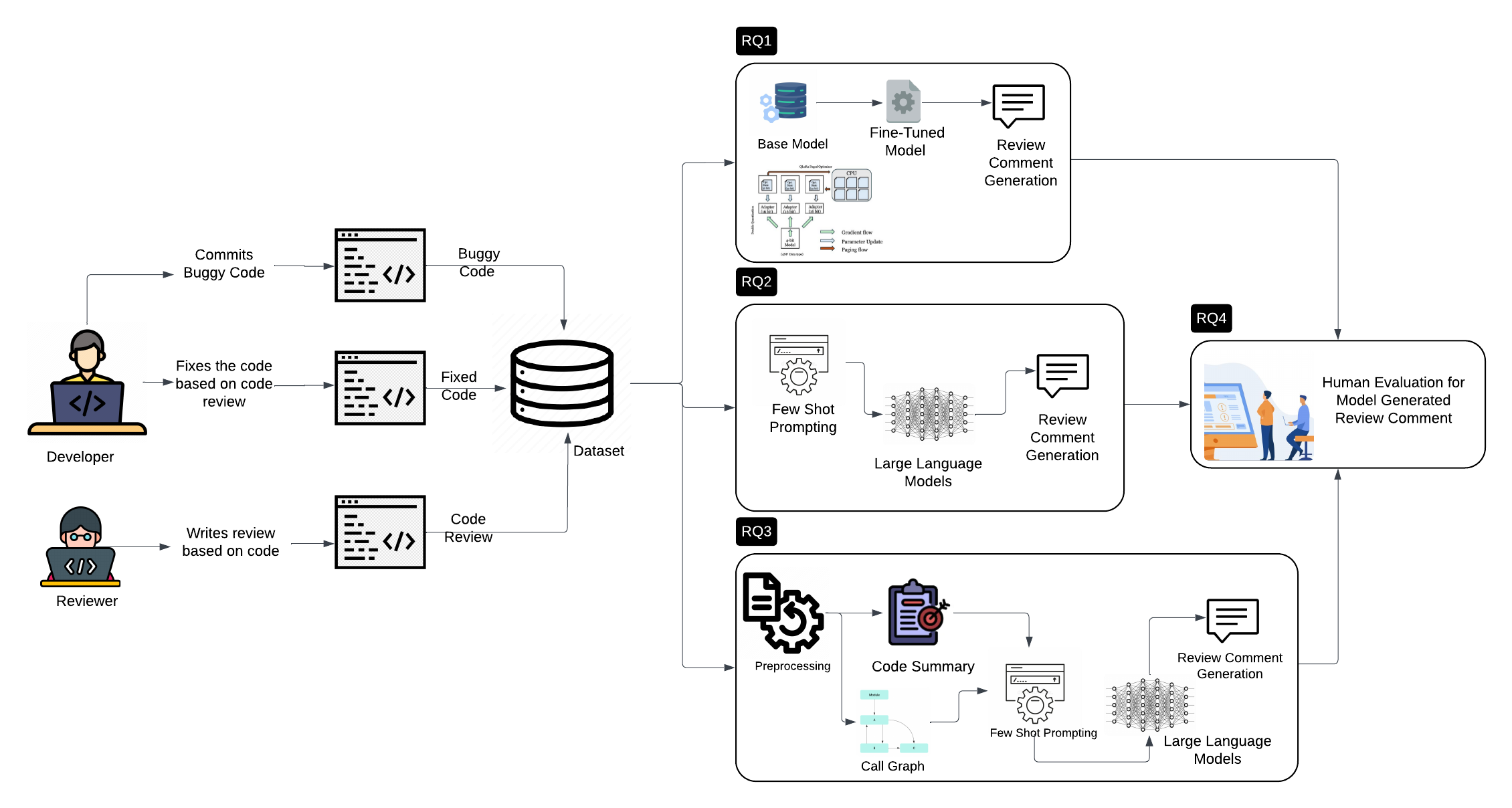}}
\caption{Overview of the methodology}
\label{fig:overview}
\end{figure*}

\section{Background}
We provide a brief overview of LLMs, few-shot prompting and parameter-efficient QLoRA fine-tuning approaches.

\subsection{Large Language Models}
Large Language Models (LLM) represent a novel class of computational models with millions and billions of trained model parameters designed to process, understand, and generate human-like text. LLMs for software engineering tasks are typically divided into two types: general-purpose models and code-focused models. General-purpose models, such as the GPT series and Gemini \cite{team2023gemini}, are designed for a broad range of tasks, including question-answering, summarizing, and generating dialogues. Code-specific LLMs, like Code Llama \cite{roziere2023code}, are specialized in programming languages and excel in generating, completing, and debugging code. Built on the foundation of the Transformer architecture with multi-headed attention for efficient sequence processing, these models have evolved significantly since 2017. Transformers replaced RNNs, enabling better long-range dependency handling and allowing models like GPT to scale to trillions of parameters and multimodal capabilities. Gemini, based on Google’s T5 \cite{Raffel2019-t5} model, has further advanced multi-modal capabilities, processing text, images, audio, and video with an extended context window, making it valuable for tasks like code summarization and refinement.

\subsection{Few-shot Prompting}
Until 2020, fine-tuning was pre-dominant for adapting the models to a specific task. However, with advancements in LLMs, prompt engineering has emerged as an efficient alternative \cite{pornprasit2024gpt, Feng2024-codeprompt1, White2023-persona-catalog, Kim2023-chainthought}. A prompt is a structured natural language input that directs Large Language Models (LLMs) to execute a specified task, incorporating instructions and contextual information to enhance output relevance of test queries. While Large Language Models can perform well with zero-shot, they achieve better performance with few-shot prompting \cite{brown2020language}. This technique introduces a limited number of labeled examples within the prompt as contextual information, illustrating the input-output relationship to guide the model’s responses. Top few-shot context suited for the test query can be retrieved from the training dataset utilizing a separate ranking algorithms.

\subsection{QLoRA : Parameter Efficient Fine-Tuning}
Modern Large Language Models have billions of parameters, hence fine tuning these models requires a vast amount of memory and high-end GPU machines. A full finetuning process can also be extremely time and energy consuming, as it involves training all the layers and parameters with task-specific data. Several parameter-efficient fine-tuning techniques, also known collectively as PEFT methods \cite{li2021prefix, lester2021power,  hu2021lora} have been proposed and applied to decrease the memory usage
and training cost while maintaining the accuracy of the fine-tuned model to a reasonable extent \cite{lu2023llama}. 
Quantized LoRA, best known as QLoRA \cite{dettmers2024qlora} is a quantized version of LoRA that introduces quantizing the transformer model to 4-bit NormalFloat (NF4) precision with double quantization processing from typical 16-bit FloatingPoint (FP16). It also utilizes a paged optimizer to deal with memory spikes seen when training with longer batches, eventually making fine-tuning possible with more limited computational resources.


\section{Methodology}

Figure \ref{fig:overview} provides a brief graphical overview of our study. We collected review comments, along with the code snippets before and after code review from the CodeReviewer \cite{li2022automating} dataset. To answer \textbf{RQ1}, we fine-tuned models from Meta Llama series (Llama 2, Code Llama, Llama 3, Llama 3.1, Llama 3.2). For addressing \textbf{RQ2} and \textbf{RQ3}, we few-shot prompted models from GPT (GPT-3.5 Turbo, GPT 4o) and Gemini (Gemini-1.0 Pro) series, and experimented with different prompt augmentation strategies. Finally, we invited a group of software developers from industry to participate in an anonymous human evaluation study designed to answer our proposed \textbf{RQ4}. The following subsections explain each of these steps in detail.

\subsection{Dataset}
We used the CodeReviewer\cite{li2022automating} dataset for all of our experiments. Introduced by Microsoft, this dataset was collected from publicly available high-quality open-source repositories. It covers nine most popular languages including C, C++, C\#, Go, Java, JavaScript, PHP, Python, and Ruby. This dataset was processed for three downstream tasks (code change quality estimation, review comment generation, and code refinement). As our experiment focuses on review comment generation task, this dataset fits well with our experiments. The dataset was already split into three parts (Training, Validation, and Test data). As data from the same project can result in biased test and validation data, the dataset was split at the project level. Hence, there is no correlation between the three parts of the dataset. We separated all the necessary fields of the dataset for our experiment, including \textit{old file} (the file before the pull request), \textit{code diff}, and a \textit{review comment}.\\

\begin{table}[htbp]
\renewcommand{\arraystretch}{1.25}
\caption{Overview of the Dataset}
\label{dataset}
\begin{center}
\begin{tabular}{c c c}
\hline
\textbf{Dataset} & \textbf{Split Type} & \textbf{Count}\\
\hline
\multirow{5}{*}{CodeReviewer} & Train Set & $\sim 118k$ \\
& Validation Set & $\sim 10k$\\
& Test Set & $\sim 10k$ \\
& Test Subset 1 & $5000$ \\
& Test Subset 2 & $500$ \\
\hline
\end{tabular}
\end{center}
\end{table}

We intended not to utilize the whole test dataset for evaluation to keep our prompting cost within limit, as closed-source LLMs like GPT-3.5 and GPT-4 series impose a significant amount of cost overhead when applied to such a huge dataset with thousands of entries. Consequently, for our experiment purpose, we randomly picked 5000 and 500 samples from the whole test set and only evaluated further prompting, fine-tuning, and ablation study using these two test subsets. Table \ref{dataset} shows the overview of the dataset used in our experiments. Wilcoxon-signed rank test conducted on these subsets with results from section \ref{result-1}, \ref{result-2} and \ref{rq3-result} suggest no statistically significant difference among these different versions of the test set.


\subsection{RQ1: Parameter Efficient Quantized Fine-tuning for RCG}

In this section, we describe in detail how we fine-tuned open-source LLMs for the review comment generation task. We utilized QLoRA, an effective parameter-efficient fine-tuning (PEFT) technique that combines together the idea of working with low-rank matrices and 4-bit quantization. The motivation for going with the PEFT approach was to ensure fine-tuning in a constrained, low-resource setting which can be achieved with a consumer grade GPU configuration.\\


\subsubsection{Data Preprocessing for Supervised Fine-tuning}

\begin{figure}[htbp]
\centerline{\includegraphics[width=0.5\textwidth]{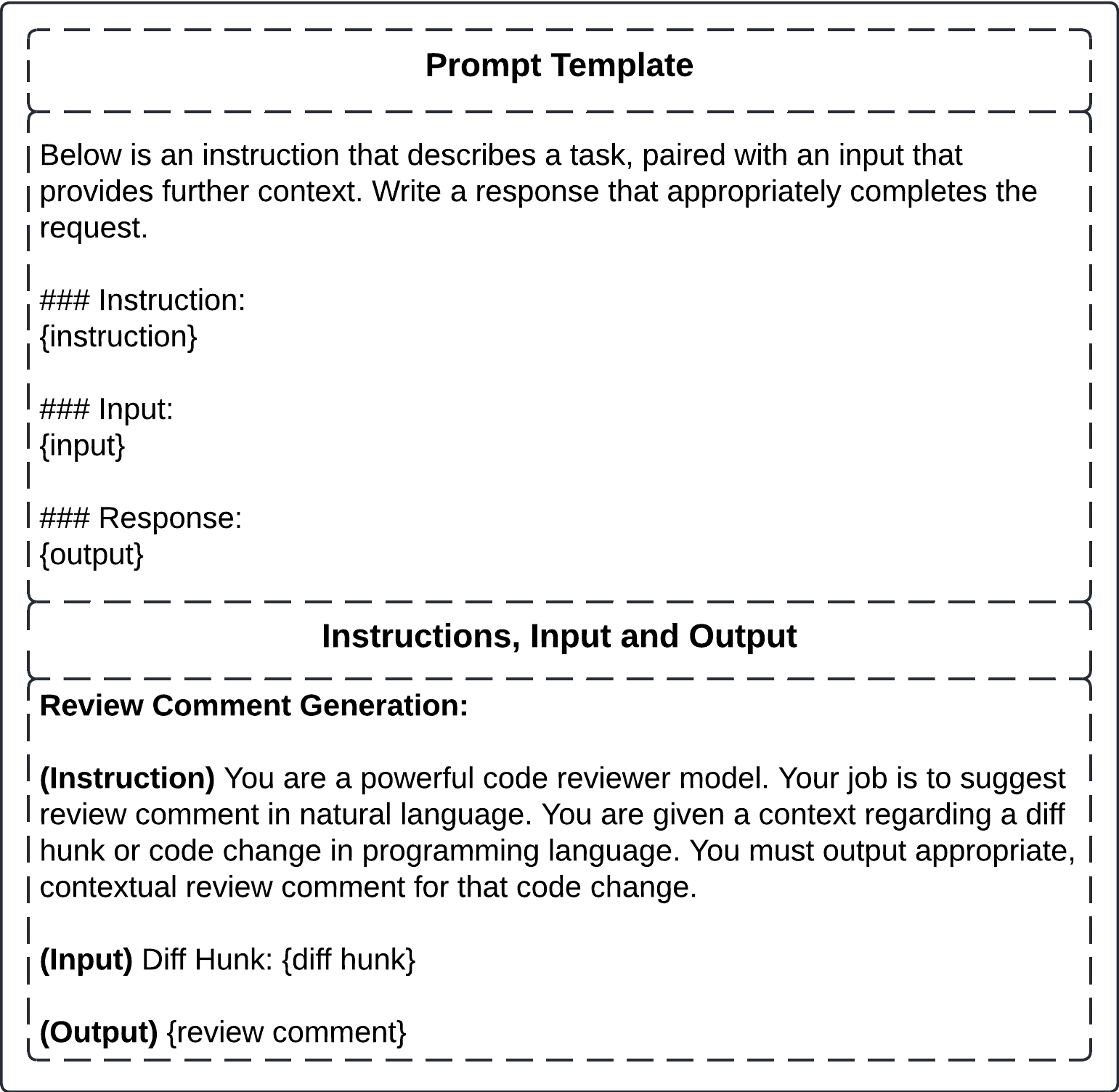}}
\caption{Prompt template for supervised fine-tuning with instruction, input, output specified}
\label{fig:finetune-prompt}
\end{figure}

LLMs, when trained to follow some specific form of instructions tend to show superior performance in practice. Fine-tuning on a diverse range of multi-task datasets with rich natural language instructions has been proven to demonstrate performance enhancement on completely unseen tasks ~\cite{wei2021finetuned, ouyang2022training}. Hence in this work, we modified the dataset to contain a suitable instruction-following prompt for the code review comment generation task. We crafted our template inspired by Stanford Alpaca~\cite{alpaca} and used the original dataset from CodeReviewer~\cite{li2022automating} to perform fine-tuning in a supervised fashion. The modified data structure follows the \{instruction, input, output\} format, similar to the framework introduced in ~\cite{wang2022self}. We show our prompt template and corresponding sample instruction, input, and output in Figure \ref{fig:finetune-prompt}.\\

\subsubsection{Fine-tuning Open-Source Models}

Using the instruction-following dataset that we prepared, we fine-tuned various generations and variants of the Llama, the leading LLM series by Meta in the open-source scenario. \textbf{Llama 2}~\cite{touvron2023llama}, the direct successor of the Meta Llama model was pretrained on 2 trillion tokens. \textbf{Code Llama}~\cite{roziere2023code} was built on top of Llama 2 after training with fill-in-the-middle capabilities on code-specific datasets. We fine-tuned the 7B base version of both models for our task. \textbf{Llama 3}\footnote{\url{https://github.com/meta-llama/llama3}} even surpassed the previous models in the Llama series as openly accessible LLMs, followed by \textbf{Llama 3.1} (with a greater window of $128k$ tokens) and \textbf{Llama 3.2} (lightweight, quantized version). We used the 3B Instruct version for our purpose.\\

\subsubsection{Experimental Setup}

We implemented our fine-tuning experiment using a couple of popular open-source tools. We used Axolotl for fine-tuning Llama 2 and Code Llama 7B variants with QLoRA adapter for 4-bit quantization. Axolotl\footnote{\url{https://github.com/OpenAccess-AI-Collective/axolotl}} is a tool designed to streamline the fine-tuning of various AI models, offering support for multiple configurations and architectures. To fine-tune all the Llama 3 models, we used Unsloth\footnote{\url{https://github.com/unslothai/unsloth}}, which offers faster training and inference speed for more recent LLMs. All experiments were conducted on NVIDIA RTX 5000 GPU machine with 16 GB VRAM. The token length limit was set to 2048. All the models were trained for 2 full epochs (except 5 epochs for Llama 3.2) using a 32-bit paged AdamW optimizer for Llama 2 and Code Llama models, while the Llama 3 series used an 8-bit AdamW optimizer instead. We set the LoRA rank to 32, the scaling factor alpha to 16, and the dropout rate to 0.05. The micro batch size was fixed to 2 and a learning rate of 0.0002 was used with 0.01 weight decay. 


\subsection{RQ2: Few-shot Prompting for RCG}
\label{rq2-method}

In this section, we discuss in detail how we applied prompt engineering to our task and dataset.\\
\subsubsection{Prompted Closed-Source Models} We chose three proprietary popular LLMs for our prompting experiments. Two of them, the GPT-3.5 Turbo \cite{brown2020language} and GPT-4o \cite{achiam2023gpt} models were accessed via OpenAI API, Gemini-1.0 Pro \cite{team2023gemini} model was provided by Google DeepMind API. These decoder-only models
generate text by predicting the next token in a sequence given the previous tokens, which makes these models particularly effective for tasks involving text generation and completion. \\

We chose the \textit{instruct} version of the GPT-3.5 Turbo, the most cost-effective model in the GPT-3.5 series. It offers a context window of 4,096 tokens and is trained on data up to September 2021. GPT-4 omni, also known as GPT-4o, is the flagship multimodal model of OpenAI with advanced reasoning performance. It is updated with knowledge up to October 2023, providing enhanced capabilities in understanding and generating diverse types of content. We finally picked Gemini-1.0 Pro from the Gemini series of advanced multi-modal models developed by Google. It was created to scale across a wide range of tasks with up to 1M input tokens, a limit greater than the other comparable models. Gemini-1.0-pro offers a context window of 32k tokens and is noted for its impressive output speed of 86.8 tokens per second, making it comparatively faster.\\

\subsubsection{Few-shot Prompt Design}


A prompt is a structured natural language input presented to a language model, designed to receive a specific response tailored to a particular task. While, in many cases, large language models show outstanding performance with zero-shot prompting (without any example), it might need some examples to be provided for complex tasks. Few-shot prompting is a widely used technique to enable in-context learning \cite{Liu2021-contextgpt, Geng2024-codeprompt3, Gao2023-codeprompt2} where we provide demonstrations in the prompt to steer the model to perform better.\\

Thoughtful, well-crafted prompts are crucial for leveraging the performance of generative models like GPT-3.5 and 4 series. Review Comment Generation is a complex task that requires much focus and contextual knowledge about the code snippet. Generated review comments need to be precise, specially on the parts where the code is to be modified. A comment needs to be informative, relevant and well-explained to denote where to fix issues in the code. In our prompt, we include \{\textit{Instruction}$_{optional}$ + \textit{Exemplars} + \textit{Query}$_{test}$\}. Below we discuss these in detail.\\

\begin{itemize}
    \item \textbf{\textit{Instruction:}}  Instructions are only added in the GPT-4o chat model, to avoid producing overly long code reviews and encourage concise review comments that are at once relevant and informative to the code changes. We took inspiration for our prompt design from \textit{Specificity and Information}, \textit{Content and Language Style}, \textit{User Interaction and Engagement} and \textit{Prompt Structure and Clarity} categories as shown in \cite{bsharat2023principled}, and \textit{Emotional Stimulus} as presented in \cite{li2023large}. Upon design experiments, we chose \textit{"Please provide formal code review for software developers in one sentence for following test case, implementing few-shot learning from examples. Don't start with code review/review. Just give the answer."} as our prompt.
    
    \item \textbf{\textit{Exemplars:}}  Few-shot ($k$-shot where $k$ can be 3, 5, 10 for example) exemplars were the examples collected from the training dataset to guide the model towards generating output with desired accuracy and format. For each sample from our test subset in our study, we employed BM-25 \cite{robertson2009probabilistic}, the popular information retrieval and ranking algorithm to retrieve the most relevant $k$ samples from the training set. Following the original CodeReviewer study \cite{li2022automating}, each example of our prompt contained the following contents:
    \begin{itemize}
        \item \textbf{Code Diff: } A code snippet showing the changes in the code, denoted as \textbf{patch} in the CodeReviewer dataset.
        \item \textbf{Code Review:} The corresponding review comment, originally the \textbf{msg} portion of the CodeReviewer dataset samples.
    \end{itemize}
    \item \textbf{\textit{Query}$_{test}$ :} For evaluating the model, we used test queries structured similarly to the training examples. Each test case of our Test Subset 1 contains only the \textbf{Code Diff}. We appended "\textbf{Code Review}" tag to indicate the model to complete the desired review comment in the prompt. 
\end{itemize}

We also added function call graph and code summary components to each of our $k$-shot samples (similarly, in the test queries) to address the RQ3, which we discuss in section \ref{rq3} in detail.

\begin{figure}[htbp]
\centerline{\includegraphics[width=0.5\textwidth]{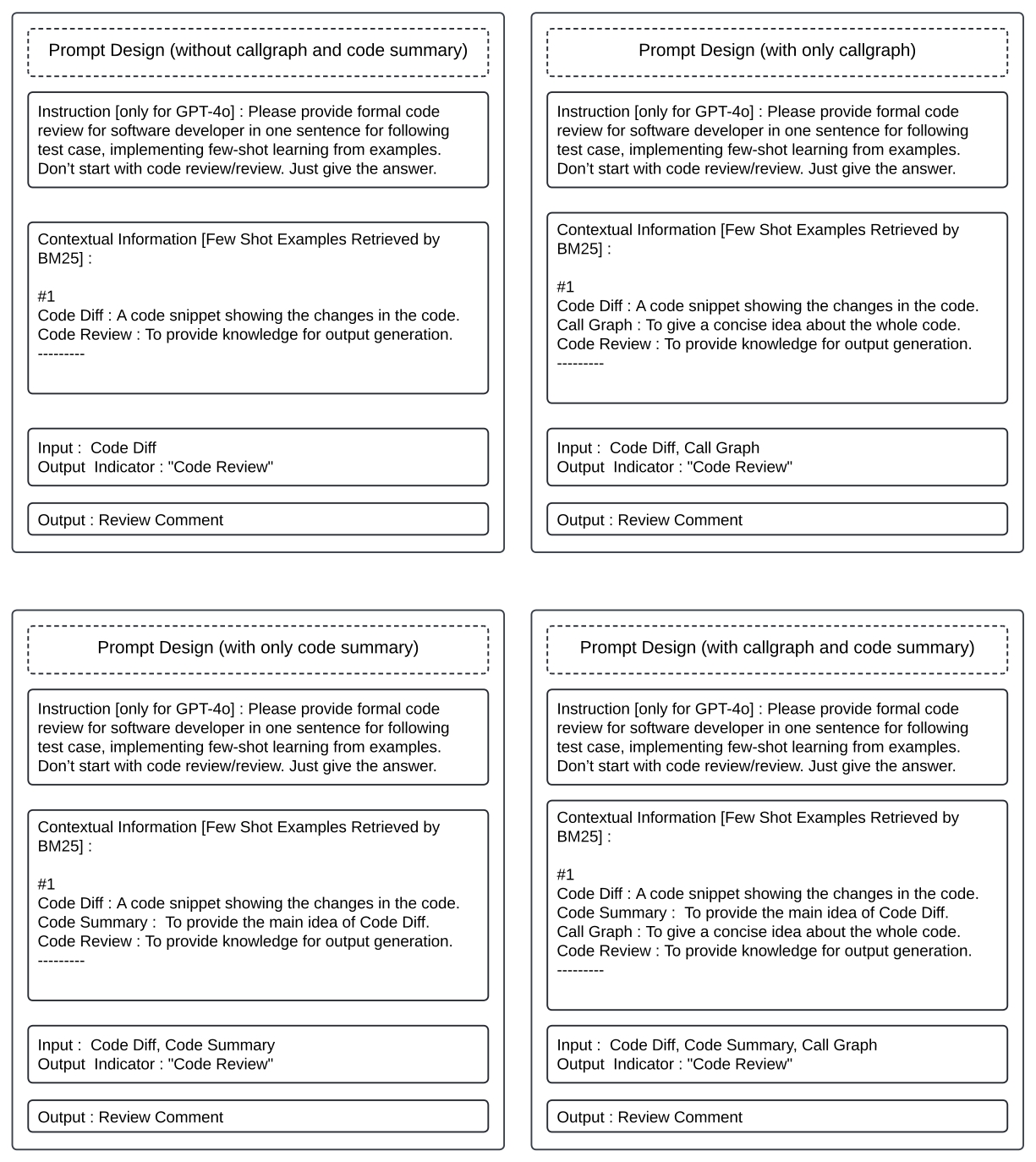}}
\caption{Prompt designs for few-shot prompting}
\label{fig:Prompt Design for Code Review Task}
\end{figure}

\subsection{RQ3: Impact of Semantic Metadata Augmented Prompts
for RCG}
\label{rq3}
In this section, we discuss the key steps of dataset preprocessing to produce function call graph and code summary. Then we present how we incorporated these into our prompting pipeline already discussed in section \ref{rq2-method}.\\

\subsubsection{Extracting Function Call Graph}
As \textbf{code diff} represents a small part of the whole code base, carrying not much information about the other parts of the code, the challenge remained: how can we enable our model to recognize the gist of the code base? To address this issue, we hypothesized that semantic data flow information, like function call graphs, is an important contributing factor to the deeper understanding of our code base. A function call graph is a control flow graph representing which function is called from other functions. It creates a directed graph where each node represents a function or module and each edge symbolizes the call from one function to another. \\

To achieve this, we extracted syntactic details from the given code sample first, leveraging Abstract Syntax Tree (AST). AST, in essence, is a data structure that captures the syntactic structure of a program or code. It forms a tree where each node denotes a construct occurring in the code. We parsed each old file code \textbf{`oldf'} from our dataset to generate an AST to identify key elements like function calls and definitions, using \textbf{Tree-sitter}\footnote{\url{https://github.com/tree-sitter/tree-sitter}}, a popular parser generator tool and incremental parsing library with support for multiple programming languages.

\begin{figure}[htbp]
\centerline{\includegraphics[width=0.5\textwidth]{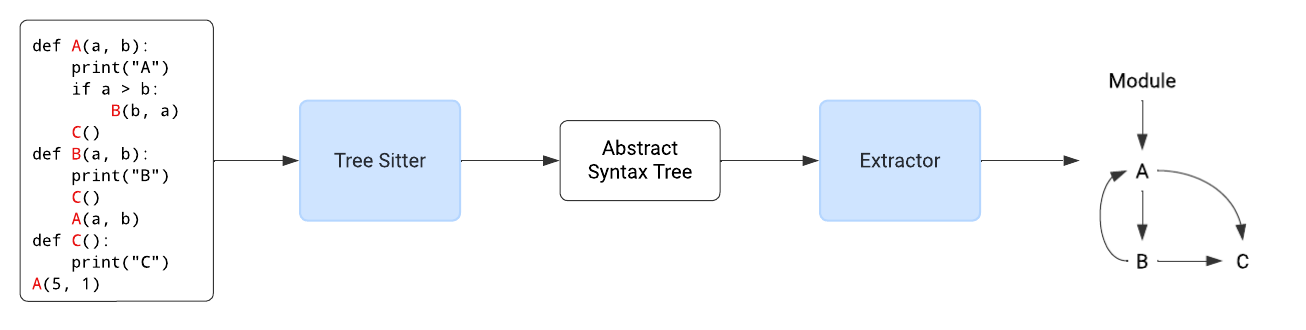}}
\caption{Call graph extraction workflow}
\label{fig:callgraph-flow}
\end{figure}

\begin{figure}[htbp]
\centerline{\includegraphics[width=0.5\textwidth]{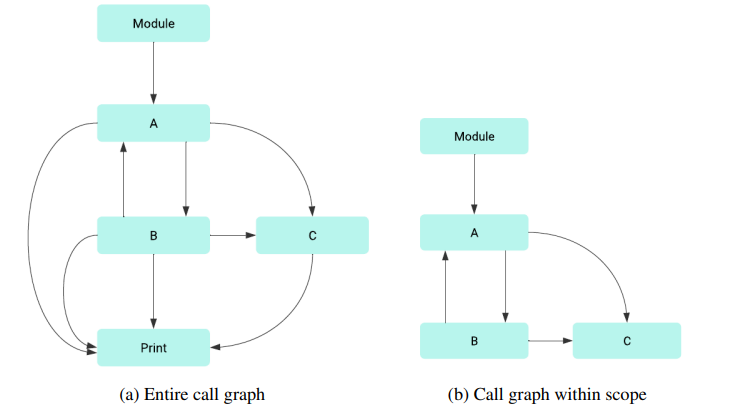}}
\caption{Scope resolution of call graphs}
\label{fig:callgraph-scope}
\end{figure}

As shown in Figure \ref{fig:callgraph-flow}, we traverse the whole tree to identify the function calls. For each of the nine languages, we built an extractor that extracts the function call relationships, from which we constructed adjacency lists to neatly represent the call graphs. Initially, we included all the function calls with scope resolution operators to distinguish different functions in different modules, but this led to excessively large call graphs, impacting performance. To address this,  we chose to remove the scope resolution operators and duplicate function calls. Finally, we excluded external (e.g., library) function calls to maintain focus on the core code structure. Figure \ref{fig:callgraph-scope} shows the overview of these modifications.

\subsubsection{Generating Code Summary}

We hypothesized that including code summarization would capture the overall context of code changes in code diffs, improving code review comment generation. Initially, we attempted to summarize the entire code, but token limitations affected the performance. We then focused on summarizing only the \textit{function} related to the code diffs.
\begin{figure}[htbp]
\centerline{\includegraphics[width=0.5\textwidth]{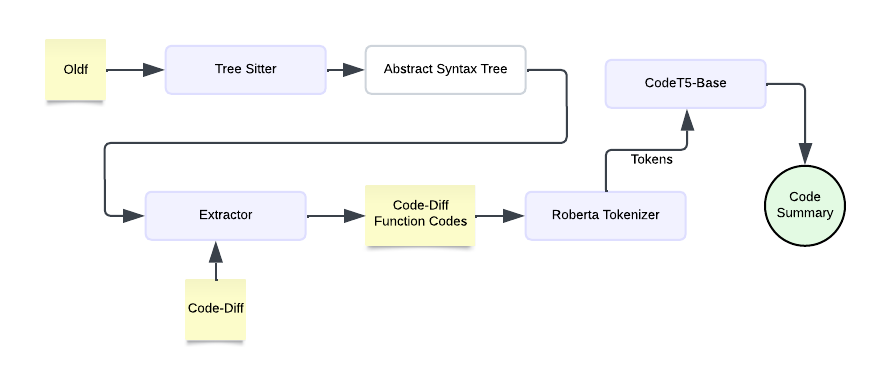}}
\caption{Code summary generation workflow}
\label{fig:/Code Summary Pipeline}
\end{figure}

Using an AST, we identified the relevant function for each code diff, building extractors for nine languages that extract the start and the end of the function and recognize language-specific
function definitions and handling edge cases such as nested functions, and class methods. If the code diff was not inside any function, we extracted the code around the code diff. We tokenized the extracted code using CodeT5’s RoBERTa-based tokenizer, splitting larger functions into smaller chunks as needed. These tokenized chunks were then fed into the \textbf{CodeT5} \cite{wang2021codet5} model, generating summaries that were appended to our prompts to enhance review accuracy.
Figure \ref{fig:/Code Summary Pipeline} outlines the key steps in this summary generation workflow.\\

\subsubsection{Experimental Setup}

Here we discuss the model hyperparameters we experimented with. We used OpenAI and Google provided APIs for invoking the experimental models. We explored different model temperatures ($temp = 0.5, 0.7$) and numbers of few-shot samples ($k = 3, 5$) to generate review comments. We finally report the results for $temp = 0.7$, and few-shot count $k = 5$ as they showed comparatively better results. We also picked top $n=5$ outputs by the models, calculated their BLEU scores, and kept the best-performing result only for further comparison. We then follow the same set of hyperparameters (except setting few-shot count $k = 3$) to prompt the Google Gemini-1.0 Pro model as well. The maximum number of tokens to be generated by the model was set to 100. We leave the rest of the available API parameters (frequency and present penalty, top\_p) to their default values. The structure of all the prompts is shown in Figure \ref{fig:Prompt Design for Code Review Task}.\\



\subsubsection{Ablation Studies}

To understand individual contributions of function call graph and code summary, we generate all combinations of prompt including without any and both of the augmentations, and augmenting with only call graph and only summary. The details are discussed in the section \ref{rq3-result}.

\subsection{RQ4: Real-World Developers Perspectives on LLM Generated Review Comments}

\subsubsection{Web portal Design}
We developed a web portal dedicated to present participants with essential components for each assessment: \textbf{code snippet}, \textbf{code summary} generated from the snippet, \textbf{ground truth} provided as a reference, and \textbf{code reviews generated by multiple models}, with model names anonymized to prevent bias. The reviewers were asked to rank the model generated comments based on three of the qualitative metrics found in literature: \textbf{relevence score, information score and explanation clarity score} \cite{li2022automating}. To avoid any misunderstanding, the instructions were presented on a separate page, ensuring that participants could review the guidelines thoroughly before proceeding with the evaluation. The backend of the portal was implemented using Node.js with the Express framework, while the frontend was developed with React.js and TypeScript to provide a responsive and interactive user interface. Participant data and study feedback were stored securely in a PostgreSQL database hosted on Render. \\

\subsubsection{Participants} We promoted the web portal among the professional software engineering community around the authors' network, as the study focuses on code review tasks. Participants were recruited on a rolling basis to ensure a diverse sample to capture a broad spectrum of programming skills and domain knowledge. Each participant was compensated for their time and feedback. We recruited 8 participants, each affiliated with reputed software industry companies in the country. Each code review example was evaluated twice, with feedback from two distinct participants, to enhance the reliability and robustness of our analysis.

\subsection{Evaluation Metrics}

We provide a brief description of the evaluation metrics used for
the review comment generation task in this section.

\begin{itemize}
    \item \textbf{BLEU} \cite{papineni2002bleu} (Bilingual Evaluation Understudy) is one of the widely acknowledged metrics to measure the text generation performance of language models. \textbf{BLEU-4} is the weighted geometric mean of all the modified 4-gram precisions, scaled by the brevity penalty. The modified n-gram precision used here adjusts for cases where the generated candidate text may have n-grams that do not perfectly match any n-grams in the reference texts.
    
    \item \textbf{BERTScore} \cite{bert-score} leverages pre-trained contextual embeddings from the BERT \cite{Devlin2018-bert} model and finds out the average F1 score, which is a harmonic mean of precision and recall, providing a single metric to assess the quality of generated sentences compared to the ground truth. It has been shown to correlate with human judgment better in terms of sentence-level similarity instead of word-level matching. A higher BERTScore indicates better similarity between the predicted sentences and the reference sentences.

    \item \textbf{Human Evaluation} metrics include the following scores:

    \begin{itemize}
        \item \textbf{Relevance score} measures how much the review comment aligns with the overall content.
        
        \item \textbf{Information score} assesses the completeness of the information provided in the review comment.
        
        \item \textbf{Explanation clarity score} evaluates the clarity of the explanations offered within the review comment.
    \end{itemize}
All these qualitative human judgment metrics were rated on a scale of 0 to 5, with 5 indicating the best score.
\end{itemize}

\begin{figure*}[htbp]
\centerline{\includegraphics[width=\textwidth]{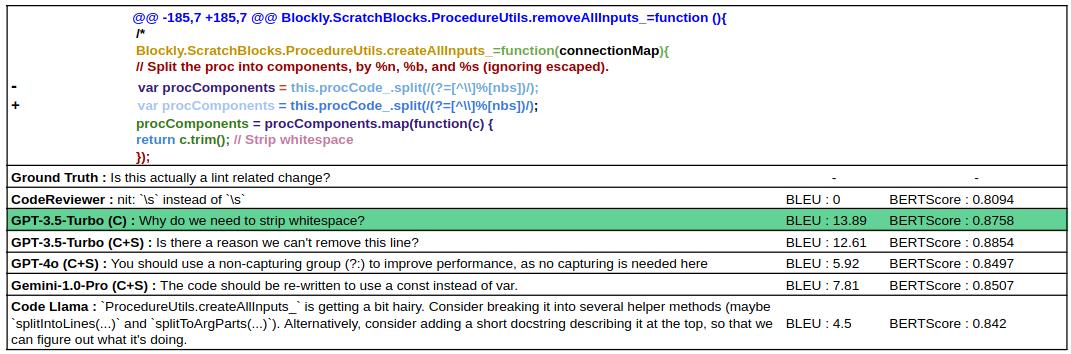}}
\caption{Model generated code review comments for a sample code diff}
\label{fig:examples}
\end{figure*}

\section{Results}


In this section, we answer each of the research questions taking evidence from our experiment results. 
\subsection{RQ1: How effective is code review comment generation using fine-tuned open-source Large Language Models?}
\label{result-1}

\begin{table}[htbp]
\renewcommand{\arraystretch}{1.25}
\caption{Comparison of fine-tuned Llama Models on Test Subset 1}
\label{fine_tune_Model_comparison}
\begin{center}
\begin{tabular}{c c c}
\hline
\textbf{Model} & \textbf{BLEU-4} & \textbf{BERTScore}\\
\hline
CodeReviewer (223M) & 4.28 $_{baseline}$ & 0.8348 $_{baseline}$ \\
\hline
Llama 2 (7B) & 5.02 $_{\text{\color{darkgreen}+17.29\%}}$ & 0.8397 $_{\text{\color{darkgreen}+0.59\%}}$ \\
Code Llama (7B) & \textbf{5.58 $_{\text{\color{darkgreen}+30.37\%}}$} & 0.8480 $_{\text{\color{darkgreen}+1.58\%}}$ \\
Llama 3 (8B) & 5.27 $_{\text{\color{darkgreen}+23.13\%}}$ & 0.8476 $_{\text{\color{darkgreen}+1.53\%}}$ \\
Llama 3.1 (8B)& 5.38 $_{\text{\color{darkgreen}+25.7\%}}$ & \textbf{0.8483 $_{\text{\color{darkgreen}+1.62\%}}$} \\
Llama 3.2 (3B) & 4.54 $_{\text{\color{darkgreen}+6.07\%}}$ & 0.8371 $_{\text{\color{darkgreen}+0.28\%}}$ \\
\hline
\end{tabular}
\label{tab1}
\end{center}
\end{table}
We present our QLoRA fine-tuned open-source models' performance on Test Subset 1 in Table \ref{fine_tune_Model_comparison}. The results demonstrate a notable improvement in both BLEU-4 and BERTScore metrics for all fine-tuned models compared to the baseline CodeReviewer model. Code Llama (7B) achieved the highest BLEU-4 score of 5.58, reflecting a 30.37\% increase over the pretrained CodeReviewer model. In terms of BERTScore, Llama 3.1 (8B) scored the highest value of 0.8483, representing a 1.62\% improvement. 

\begin{findingsbox}
    \textbf{RQ1 Findings:} Supervised fine-tuned (QLoRA) models showed notable performance in review comment generation task. All open-source LLMs tested with this approach (Llama 2, Code Llama and Llama 3 variants) outperformed the baseline. Specially Code Llama, which is designed for processing code, demonstrates a particularly significant improvement. 
\end{findingsbox}

\subsection{RQ2: How well do the closed-source Large Language Models perform in code review comment generation task when prompt engineered in a few-shot setting?}
\label{result-2}

Next, we used few-shot prompting for review comment generation using leading closed-source LLMs. Due to the incurring API cost of the closed-source models, we conducted our initial experiment on the smaller Test Subset 2. We report the promising results of 5-shot prompting in Table \ref{tab3}. As it indicates, our experimental models outperform the baseline even without any further data augmentation.

\begin{table}[htbp]
\renewcommand{\arraystretch}{1.25}
\caption{Comparison of few-shot prompted closed-source LLMs on Test Subset 2}
\label{tab3}
\begin{center}
\begin{tabular}{c c c}
\hline
\textbf{Model} & \textbf{BLEU-4} & \textbf{BERTScore}\\
\hline
CodeReviewer & 4.28 $_{baseline}$ & 0.8348 $_{baseline}$ \\
\hline
GPT-3.5 Turbo & \textbf{8.13 $_{\text{\color{darkgreen}+89.95\%}}$} & \textbf{0.8509 $_{\text{\color{darkgreen}+1.93\%}}$} \\
Gemini-1.0 Pro & 7.85 $_{\text{\color{darkgreen}+83.41\%}}$ & 0.8509 $_{\text{\color{darkgreen}+1.93\%}}$ \\
GPT-4o & 6.92 $_{\text{\color{darkgreen}+61.68\%}}$ & 0.8505 $_{\text{\color{darkgreen}+1.88\%}}$ \\
\hline
\end{tabular}
\end{center}
\end{table}

GPT-3.5 Turbo achieved an impressive performance improvement of $89.95\%$ with a BLEU-4 score of 8.13 surpassing other experimental models. Gemini-1.0 Pro performed second best, with a BLEU-4 score of 7.85 (and $83.41\%$ boost over the baseline). GPT-4o model performed comparatively poor, scoring a 6.92 BLEU-4 (with $61.68\%$ improvement over the baseline). On BERTScore metric, GPT-3.5 Turbo and Gemini-1.0 Pro achieved 0.8509, both gaining a performance upgrade of $1.93\%$. 

\begin{findingsbox}
    \textbf{RQ2 Findings:} Closed-source LLMs proved to be highly effective, as they improved over the baseline, exploiting the effectiveness of few-shot prompting for review comment generation task. Among the models tested, GPT-3.5 Turbo achieved the most significant performance gain. Gemini-1.0 Pro and GPT-4o also showed impressive results.
\end{findingsbox}

\subsection{RQ3: When incorporated into prompts, what are the impacts of function call graph and code summary in improving review comment generation performance?}
\label{rq3-result}

\begin{table}[htbp]
\renewcommand{\arraystretch}{1.25}
\caption{Prompting closed-source LLMs on Test Subset 2 with augmented Function call graph and code summary}
\begin{center}
\begin{tabular}{c c c}
\hline
\textbf{Model} & \textbf{BLEU-4} & \textbf{BERTScore}\\
\hline
CodeReviewer & 4.28 $_{baseline}$ & 0.8348 $_{baseline}$ \\
\hline
GPT-3.5 Turbo ($W$) & \textbf{8.13 $_{\text{\color{darkgreen}+89.95\%}}$ $_{(ref)}$} & 0.8509 $_{\text{\color{darkgreen}+1.93\%}}$ $_{(ref)}$ \\
GPT-3.5 Turbo ($C+S$) & 7.83 $_{\text{\color{darkgreen}+82.94\%}}$ $_{\text{\color{red}(-3.69\%)}}$ & 0.8514 $_{\text{\color{darkgreen}+1.99\%}}$ $_{\text{\color{darkgreen}(+0.06\%)}}$ \\
\hline
Gemini-1.0 Pro ($W$) & 7.85 $_{\text{\color{darkgreen}+83.41\%}}$ $_{(ref)}$& 0.8509 $_{\text{\color{darkgreen}+1.93\%}}$ $_{(ref)}$
\\
Gemini-1.0 Pro ($C+S$) & 7.77 $_{\text{\color{darkgreen}+81.54\%}}$ $_{\text{\color{red}(-1.02\%)}}$& 0.8496 $_{\text{\color{darkgreen}+1.77\%}}$$_{\text{\color{red}(-0.15\%)}}$\\
\hline
GPT-4o ($W$) & 6.92 $_{\text{\color{darkgreen}+61.68\%}}$ $_{(ref)}$& 0.8505 $_{\text{\color{darkgreen}+1.88\%}}$ $_{(ref)}$ \\
GPT-4o ($C+S$) & 7.39 $_{\text{\color{darkgreen}+72.66\%}}$ \textbf{$_{\text{\color{darkgreen}(+6.79\%)}}$}& 
 \textbf{0.8528 $_{\text{\color{darkgreen}+2.16\%}}$ $_{\text{\color{darkgreen}(+0.27\%)}}$}\\
\hline
\end{tabular}
\label{tab4}
\end{center}
\end{table}

\begin{table}[htbp]
\renewcommand{\arraystretch}{1.25}
\caption{Further Ablation Study on Function call graph and code summary with GPT-3.5 Turbo on Test Subset 1}
\begin{center}
\begin{tabular}{c c c}
\hline
\textbf{Model} & \textbf{BLEU-4} & \textbf{BERTScore}\\
\hline
GPT-3.5 Turbo ($W$) & 8.32 $_{reference}$ & 0.8519 $_{reference}$ \\ 
GPT-3.5 Turbo ($C$) & \textbf{8.36 $_{\text{\color{darkgreen}+0.48\%}}$} & \textbf{0.8523 $_{\text{\color{darkgreen}+0.05\%}}$} \\
GPT-3.5 Turbo ($S$) & 8.20 $_{\text{\color{red}-1.44\%}}$ & 0.8518 $_{\text{\color{red}-0.01\%}}$ \\
GPT-3.5 Turbo ($C+S$) & 8.27 $_{\text{\color{red}-0.60\%}}$ & 0.8515 $_{\text{\color{red}-0.05\%}}$ \\
\hline
\end{tabular}
\label{tab5}
\end{center}
\end{table}

In this section, We report the effect of appending both function call graph and code summary into prompts in Table \ref{tab4}. We used the smaller Test Subset 2 of the dataset to evaluate our experimental closed-source LLMs with this proposed modification. In terms of both metrics, call graph and summary augmentation failed to guide the GPT-3.5 Turbo and Gemini-1.0 Pro models in the right direction, where these models have a context window of 4k and 32k tokens respectively. Manual inspection revealed that GPT-3.5 Turbo mostly got affected due to the large number of input tokens in the prompt after the proposed augmentation. On the other hand, GPT-4o with its context window of 128k tokens performs noticeably better when augmented with the same metadata, as it could handle much bigger input contexts. GPT-4o improved 6.79\% BLEU-4 score and 0.27\% BERTScore relative to the version before augmentation. However, even with such a relative improvement, GPT-4o falls short behind GPT-3.5 Turbo performance.\\

Contrary to our hypothesis, incorporating both the function call graph and code summary together negatively affected the performance of a few models. So we experimented with all possible combinations of these augmentation to identify which additional information was causing a degradation in performance. To keep the API expense within limit, we used the larger Test Subset 1 to conduct this ablation study on our best performing model GPT-3.5 Turbo. Evidence from Table \ref{tab5} suggests adding standalone function call graph is advantageous while adding code summary with prompts has a negative impact on the performance. With only the function call graph added, GPT-3.5 Turbo achieves the highest BLEU-4 score of 8.36 which is over 90\% of baseline performance. Figure \ref{fig:examples} presents a test sample with multiple LLM generated review comments.


\begin{findingsbox}
    \textbf{RQ3 Findings:} Albeit being the inferior performer, GPT-4o with longer context window shows relative performance improvement when both call graph and summary were added simultaneously. Further ablation experiments suggest that, while function call graph guides the model to generate better code review, the code summary mostly affects the result negatively. Thus, when both of them are incorporated, already well-performing models like GPT-3.5 Turbo demonstrate slight performance degradation. 
\end{findingsbox}

\subsection{RQ4: How effective Large Language Models are in generating review comments from a real-world developer perspective?}
\begin{figure}[htbp]
\centerline{\includegraphics[width=0.5\textwidth]{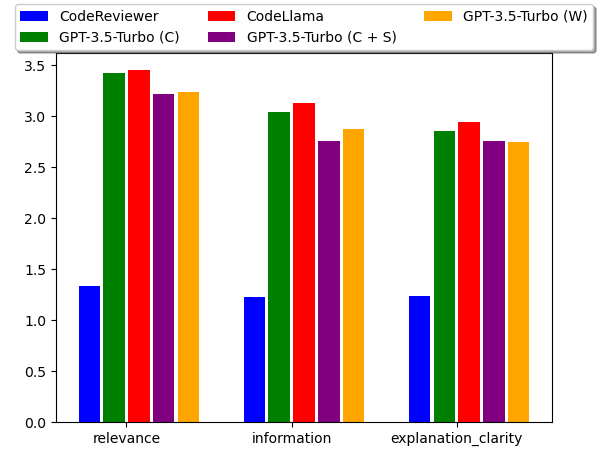}}
\caption{Comparison of LLM-generated code reviews across their average qualitative scores on a scale of 5, as perceived by developers}
\label{human-avg}
\end{figure}


\begin{table}[htbp]
\renewcommand{\arraystretch}{1.25}
\caption{Evaluation of LLM-generated code reviews by real-world developers across their average qualitative scores on a scale of 5}
\begin{center}
\begin{tabular}{c c c c}
\hline
\textbf{Model} & \textbf{Relevance} & \textbf{Information} & \textbf{Explanation Clarity} \\
\hline
Codereviewer & 1.34 & 1.22 & 1.23 \\ 
\textbf{Code Llama} & \textbf{3.45} & \textbf{3.13} & \textbf{2.95}\\
GPT-3.5-Turbo ($W$) & 3.23 & 2.87 & 2.74\\
GPT-3.5-Turbo ($C$) & 3.42 & 3.04 & 2.89\\
GPT-3.5-Turbo ($C + S$) & 3.22 & 2.76 & 2.76 \\
\hline
\end{tabular}
\label{tab6}
\end{center}
\end{table}

Further analysis of the qualitative metric scores collected from the developer study supports our previous findings that experimental models significantly outperformed the baseline. Figure \ref{human-avg} illustrates the top models' scores across relevance, informativeness, and explanation clarity metrics. The exact scores are reported in Table \ref{tab6}. Among the GPT series, \textbf{GPT-3.5 Turbo} with \textbf{Callgraph} achieved notable performance, surpassing the other two GPT-3.5 variants. Interestingly, \textbf{Code Llama} demonstrated the strongest qualitative results, possibly due to its pretraining, focused on meticulously designed code-related tasks.


\begin{findingsbox}
    \textbf{RQ4 Findings:} According to the reported experience of 8 software practitioners, LLM generated review comments showed notable performance compared to the baseline Codereviewer. Fine-tuned Code Llama has outperformed all other models in all three qualitative metrics closely followed by GPT-3.5 Turbo augmented with callgraph.
\end{findingsbox}

\section{Discussion}
\subsubsection{Observations} Although all our trained models surpassed baseline pretrained CodeReviewer (which exhibits high confidence in incorrect answers), they each have specific limitations. Although GPT-3.5-turbo is cost-effective, they can get distracted due to the limited context window size. In contrast, GPT-4o shows improved performance with a longer context window, allowing for greater focus on call graph and code summary for generating code reviews. However, due to budget constraints, we were unable to fully explore the performance of GPT-4o on the whole dataset.  On the other hand, fine-tuned Code Llama tend to generalize poor, often failing to address specific changes in greater detail.\\

\subsubsection{Future directions}
We are incorporating our model-generated code reviews into code repair tasks, experimenting with different prompt and fine-tuning strategies to improve the ability of the model to suggest effective code fixes after review.
\section{Threats to Validity}

\textbf{Threats to internal validity} are related to how the roles played by the model architecture and the configuration of hyper-parameters might impact the experiments. Due to cost and resource constraints, we explored hyper-parameters prioritizing their expected impact on model performance, while leaving others less explored. As a result, it is possible that further tuning could yield additional improvements.\\

\textbf{Threats to construct validity} include our use of the widely popular BLEU metric, as it was used in the original baseline study and relevant literature. However, its ability to reflect true performance remains uncertain. The human evaluation results indicate that Code Llama outperforms GPT-3.5 Turbo, with both models surpassing the baseline. This raises questions about the overall reliability of BLEU as a performance measure, suggesting that this might be a threat to construct validity.\\

\textbf{Threats to conclusion validity} highlight the fact that the exact same responses might not be generated by LLMs, given their inherent nondeterministic nature \cite{ouyang2023llm}. Additionally, by setting the temperature parameter to 0.7, we encourage more variability in the model's outputs. This nondeterminism in LLM can potentially undermine the conclusion validity drawn from their responses.\\

\textbf{Threats to external validity} are primarily related to the dataset used in this study. All experiments were conducted using the CodeReviewer \cite{li2022automating} dataset of Microsoft Research. As the dataset is derived from publicly available open-source repositories rather than industrial projects, the generalizability of our findings to industrial applications may be limited.
\section{Related Works}

\subsection{Automation of Code Review Activities}

There has been considerable interest in reducing the manual labor involved in code review activities. Researchers worked on code quality assessment \cite{li2022automating, shi2019automatic, gauthier2021historical, hijazi2022quality, hellendoorn2021towards}, probable reviewer recommendation  \cite{asthana2019whodo, chueshev2020expanding, pandya2022corms, rong2022modeling, sulun2021rstrace+, rebai2020multi}, suggesting review comments \cite{li2022automating, tufano2022using, balachandran2013reducing} and refining problematic code snippets \cite{li2022automating,tufano2022using, tufano2021towards, fu2022vulrepair, thongtanunam2022autotransform}. Our study focuses on the pipeline suggested by Li et al. \cite{li2022automating}. Retrieval-based approaches were initially adopted for review comment suggestion tasks. DeepMem \cite{Gupta2018-dc}, an LSTM-based model was introduced first to achieve this. Attention mechanism was added on top of LSTM architecture later on by Siow et al. \cite{Siow2020-df}. Tufano et al. \cite{tufano2022using} presented the first contribution of leveraging deep learning for this task via pretraining on code and technical language at a large scale. To improve the results, CodeReviewer \cite{li2022automating} and AUGER \cite{Li2022-ol} proposed code review-specific pretrained models. CommentFinder \cite{Hong2022-ct} showed an efficient retrieval-oriented alternative as well.\\


However, approaches so far have not considered code diffs in their pipeline. D-ACT \cite{Pornprasit2023-kn} was the first method to introduce diff-awareness in code refinement to boost performance in little differences between code base and initial commits, although they did not leverage code review comments to achieve this. CodeReviewer \cite{li2022automating}, pretrained on 9 popular programming languages, was tailored for code review activities and considered both diff-awareness and review comments altogether. Significant progress in code review activities has been possible further for the usage of unified Large Language Models, as the model size and training data continued to grow in recent times. Llama-Reviewer \cite{lu2023llama} was introduced to fine-tune the open-source Llama model tailored for code review tasks that could achieve impressive performance improvement using parameter-efficient fine-tuning approaches. It used the LoRA \cite{hu2021lora} method for fine-tuning, while there is room for improvement using the quantized counter-part \cite{dettmers2024qlora}. Additionally, more recent and better models in the Meta Llama series have been released since then, including the general-purpose model Llama 2 \cite{touvron2023llama} and Code Llama \cite{roziere2023code}, a model specifically fine-tuned on code-specific datasets. Finally, Llama 3 was also released for the open-source community.

\subsection{Pretrained and Fine-Tuned Language Models in Software Engineering}
Deep learning techniques, motivated by their impact and success in Natural Language Processing domains, have also been widely adopted in software engineering tasks. Encoder-only models like BERT \cite{Devlin2018-bert}, encoder-decoder models like BART \cite{Lewis2019-bart} and T5 \cite{Raffel2019-t5}, and decoder-only models like GPT \cite{brown2020language} are some of the notable progresses here. Pretrained code models can learn code representations with various code-specific properties including lexical, semantic, and syntactic information. Fine-tuning can help these models to be suitable for specific tasks by updating the pretrained model weights with task-specific data. Inspired by the models above, researchers proposed to further fine-tune these models on several downstream tasks in software engineering. CodeBERT \cite{feng2020codebert} and GraphCodeBERT \cite{guo2020graphcodebert} are bi-directional transformer models specifically pretrained on NL-PL pairs in 6 programming languages, with the latter introducing the incorporation of source code data flow graphs inside the model token level. These models show superior performance in tasks including natural language code search, code summarization, code clone detection, and documentation generation from code.\\ 

On the other hand, decoder-only transformer models like CodeGPT \cite{Lu2021-codegpt}, Codex \cite{Chen2021-codex} focused on generative tasks like code completion and code generation. Models like PLBART \cite{Ahmad2021-plbart} and CodeT5 \cite{wang2021codet5} can be applied for both code understanding and generation tasks. Despite covering a wide variety of code-related tasks, the models above did not pay any attention to code review activities when proposed. Tufano et al. \cite{tufano2022using} first proposed TufanoT5 that utilizes the pretrained T5 model for automating code review tasks, with CodeReviewer \cite{li2022automating} improving upon that by introducing the integration of code changes into the pretraining phase. Recently, there have been many large language models for code, including open-source community-developed LLMs like Code Llama \cite{roziere2023code}, StarCoder \cite{Li2023-starcoder}, MagiCoder \cite{Wei2023-magicoder} along with proprietary models like GPT-3.5 and GPT-4 \cite{achiam2023gpt}.

\subsection{Prompting in Software Engineering}

Prompt engineering is a good enough alternative to heavy fine-tuning that requires supervised datasets and computational resources. Providing prompts to pretrained LLMs is found to be beneficial in many code-related tasks as shown in \cite{Feng2024-codeprompt1, Gao2023-codeprompt2, Geng2024-codeprompt3, Khan2022-codeprompt4} covering code summarization, bug fixation and documentation generation. Different prompting strategies include zero-shot learning, few-shot learning \cite{brown2020language, Liu2021-contextgpt}, chain-of-thought \cite{Kim2023-chainthought}, persona \cite{White2023-persona-catalog}, etc. Not all prompting strategies are suitable for review-related activities as some of these are specifically designed for mathematical and logical reasoning tasks. Zero-shot, few-shot, and personal prompting are instruction-based and hence most suitable for code review tasks. Guo et al. \cite{Guo2024-gptzeroshot} conducted an empirical study to investigate the potential of the GPT-3.5 model for code review automation, while few-shot prompting on LLMs is still of great interest to explore. Ahmed et al. investigated prompting performance on LLMs for code summarization tasks \cite{Ahmed2022-summary}, often with augmentation of semantic metadata in prompts for GPT models \cite{Ahmed2024-asap}. We investigate the augmentation of such semantic information in a concise way into LLM prompts inspired by their work.

\section{Conclusion}
In this study, we aim to automate code review comment generation, using Large Language Models (LLMs) through carefully designed prompts with few-shot learning and parameter-efficient fine-tuning. Our prompting approach utilizes the baseline CodeReviewer dataset for our task, incorporating augmented code summaries and function call graphs after thorough pre-processing. Additionally, we applied QLoRA technique to fine-tune the open-source Large Language Models. 
Experimental results demonstrate that our strategies significantly outperform the baseline CodeReviewer in generating review comments. Furthermore, a human evaluation experiment conducted on professional developers through our designed study indicates that both finetuned code-specific LLMs and general purpose LLMs incorporating call graphs into few-shot prompts improve the quality of generated review comments, thereby further validating the effectiveness of our approach.

\bibliography{main}

\end{document}